\documentclass[journal]{rmaa}
\usepackage[applemac]{inputenc}
\usepackage{float}
\usepackage{paralist}
\usepackage{psfrag,color}
\usepackage{epsfig}
\usepackage{graphicx}

\title{Galactic kinematics of Planetary Nebulae with [WC] central star\thanks{Based on observations collected at the Observatorio Astron\'omico Nacional, SPM. B.C., M\'exico.}\thanks{Based on data obtained at Las Campanas Observatory, Carnegie Institution, Chile.}}
\author{
Miriam Pe\~na,$^1$
Jackeline S. Rechy-Garc\'ia,$^1$ 
and Jorge Garc{\'\i}a-Rojas$^{2,3}$ \\
$^1$ Instituto de Astronom{\'\i}a, Universidad Nacional Aut\'onoma de M\'exico, M\'exico.\\
$^2$ Instituto de Astrof{\'\i}sica de Canarias, Spain.\\
$^3$ Departamento de Astrof{\'\i}sica, Universidad de La Laguna,  Spain}

\shortauthor{Pe\~na, Rechy-Garc{\'\i}a \& Garc{\'\i}a-Rojas}
\shorttitle{Galactic kinematics of [WR]PNe}

\fulladdresses{
\item M. Pe\~na and J. S. Rechy-Garc{\'\i}a: Instituto de Astronom{\'\i}a, Universidad Nacional Aut\'onoma de M\'exico. Apdo. Postal 70264, M\'exico D.F., 04510, M\'exico  (miriam, jrechy@astro.unam.mx).
\item J. Garc{\'\i}a-Rojas: Instituto de Astrof{\'\i}sica de Canarias, V{\'\i}a L\'actea s/n, E-38200, La Laguna, Tenerife,  and Departamento de Astrof{\'\i}sica, Universidad de la Laguna, E-38205, La Laguna, Tenerife,  Spain.  (jogarcia@iac.es).}

\abstract{High resolution spectra are used to analyze the galactic kinematics and  distribution of a sample of  planetary nebulae with [WR] and 'wel' central star ([WR]PN and WLPN). The circular and peculiar velocities, (V$_{\rm pec}$), of the objects were derived.  The results are: a) [WR]PNe are distributed mainly in the galactic disk and they are more concentrated in a thinner disk than WLPNe and normal PNe, which corresponds to a younger population. b) The sample was separated in Peimbert's types, and it is found that  Type I PNe have  V$_{\rm pec}\leq$50 km s$^{-1}$, indicating young objects. Most of the [WR]PNe are of Type II showing V$_{\rm pec}\leq$60 km s$^{-1}$, although a small percentage is of Type III, with larger V$_{\rm pec}$  showing  that the Wolf-Rayet phenomenon in central stars can occur at any stellar mass  and in old objects. None of our WLPNe is Type I. Thus, [WR]PNe and WLPNe are unrelated objects.}

\resumen{Usando espectros de alta resoluci\'on, analizamos la distribuci\'on  y cinem\'atica gal\'actica  de nebulosas planetarias  con estrella central [WR] y 'wels' ([WR]PN and WLPN). Se calcularon  las velocidades circulares y  peculiares (V$_{\rm pec}$) de los objetos, obteniendo que: a) Las [WR]PNe pertenecen al disco gal\'actico y est\'an m\'as concentradas que las WLPNe y PN normales, lo que corresponde a objetos m\'as j\'ovenes.  b) Clasificamos la muestra en  Tipos de Peimbert encontrando que las [WR]PN de Tipo I  tienen V$_{\rm pec}\leq$50 km s$^{-1}$, como corresponde a objetos j\'ovenes; la mayor{\'\i}a de las [WR]PN son del Tipo II con V$_{\rm pec}\leq$60 km s$^{-1}$, aunque un porcentaje de la muestra es del Tipo III con V$_{\rm pec}$ m\'as alta,  indicando que el fen\'omeno  Wolf-Rayet en estrellas centrales puede ocurrir con cualquier masa estelar y en objetos viejos. Ninguna WLPN de la muestra es Tipo I. As{\'\i}, [WR]PNe y WLPNe no est\'an relacionados.}

\addkeyword{planetary nebulae}
\addkeyword{stars: Wolf-Rayet}
\addkeyword{stars: kinematics and dynamics}

\begin{document}
\maketitle

\section{Introduction}

Planetary nebulae (PNe) are formed from highly evolved low-medium mass stars, which are in the pre- white dwarf stage. The age of PN central stars fluctuates between 0.1 to about 9 Gy (Allen et al. 1998). The chemical abundances in the nebulae are typical of the moment  when the star was born, except for some elements like He, N, C, and possibly O, which have been processed in the stellar nucleus and have been partially dredged up to the surface through several dredge-up events. Thus, the youngest PNe show O, Ne, Ar, S and other $\alpha$-element abundances similar to those of the present interestellar medium, while the older objects show abundances typical of an older stellar population. Also the galactic kinematics is different in the sense that the young objects belong to the thin disk, while the oldest PNe belong to the galactic halo and appear as high velocity objects. In general most of the PNe  are  disk intermediate population.  See Peimbert (1978; 1990) for a thorough review on these subjects. 

 Considering the above, Peimbert  (1978) classified the galactic PNe in four types, according to their chemical composition and kinematics (see also Peimbert \& Serrano 1980, and Peimbert \& Torres-Peimbert 1983).  According to Peimbert (1990), who presents a more refined classification, the main characteristics of the different types are: Type I PNe are He- and N-rich objects (He/H$\geq$0.125, N/O$\geq$0.5), the initial masses of their central stars are in the range 2--8 M$_\odot$, the nebulae have,  in general, bipolar morphologies, and they belong to the young population; Type II PNe, representing the majority of the known PN sample,  are intermediate population, they have no particular He and N enrichment, the initial masses of their central stars are smaller than 2 M$_\odot$ and they show peculiar velocities lower than 60 km s$^{-1}$; Type III PNe are similar to Type II's, but their peculiar velocities are larger than 60 km s$^{-1}$ and they probably have distances to the galactic plane larger than 1 kpc. Finally, Type IV PNe are defined as those extreme population II objects that belong to the galactic halo; there are only a few objects in this category and they show very low metallicities and high peculiar velocities (see e.g., Howard et al. 1997). Peimbert's classification  has been revised by several authors. It is worth to mention for instance, that  Kingsburgh \& Barlow (1994) proposed as Type I those object with N/O $\geq$0.8. In \S 4 we will use some of the criteria given above, for classifying our objects. 
\par
Among central stars of PNe there is a particular group  which presents important atmospheric instabilities and  large mass losses. Their spectra are similar to the ones shown by massive Wolf-Rayet stars of the C series and are classified in a similar way but with the nomenclature [WC] or [WO]. PNe with this type of star represent no more than 15\% of the known sample and in the following we call them [WR]PNe. The [WR] central stars are H-deficient and their atmospheres show He, C and O, composition typical of the zone where nucleosynthesis took place (e.g., Koesterke 2001). Many works have been devoted to analyze these special type of central stars and their surrounding nebulae (e.g., G\'orny \& Stasi\'nska 1995; Pe\~na et al. 2001; Gesicki et al. 2006;  G\'orny et al. 2009; Garc{\'\i}a-Rojas et al. 2009, 2012; DePew et al. 2011, among others).

In this paper we analyze  high spectral resolution data obtained with the 2.1-m telescope and the Echelle REOSC spectrograph of the Observatorio Astron\'omico Nacional San Pedro M\'artir (OAN-SPM),  M\'exico, and with the 6.5-m Clay Telescope  equipped with the double echelle spectrograph MIKE,  at Las Campanas Observatory (LCO), Chile,
in order to study the galactic kinematical behavior  of near a hundred PNe. Of these,   a significant number  are [WR]PNe 
and the rest are ionized by normal central stars or by weak emission lines stars (wels)

 In Section 2 we present the PN sample, the observations  and the Galactic distribution of the objects; in \S 3, the heliocentric, circular and peculiar velocities are calculated for the sample with available distances. The distribution of objects in the different Peimbert Types is presented and discussed in \S 4, and our conclusions can be found in \S 5.

\section{The sample: Observations and data analysis and the Galactic distribution of objects.}
 The log of our observations, for  the whole sample,  is presented in Table 1, where we list the observatory and the observing date for each object. The sample collected at the OAN-SPM consists of 56 PNe observed from 1995 to 2001, while the sample from LCO (Clay telescope) consists of 25 objects (9 are in common with the OAN-SPM sample), observed during runs in 2006, 2009 and 2010.

For the SPM objects, the  echelle spectrograph REOSC was used at high resolution (Levine \& Chakrabarty 1994). The observed wavelength range covers from about 3600 to 6900 \AA. The description of the observations as well as the data reduction procedure can be found in  Pe\~na et al. (2001) and Medina et al. (2006).  In these works, the data of a sample of objects from Table 1 (mainly [WR]PNE),  were used  to derive and analyze the physical conditions, chemical abundances, and expansion velocities of the nebulae.  In this paper we use the data presented in those works (notice that several nebulae were observed more than once, and here we are using only one spectrum for radial velocities measurements), together with data for other objects (processed in the same way) to derive radial velocities of the nebulae. The spectral resolution of these spectra is 0.2 to 0.3 \AA~per pix which allows to determine radial velocities with a precision of about 12 to 19 km s$^{-1}$.

\begin{table*}
\caption{ Log of Observations$^{(1)}$}\label{log-observations}\centering
\begin{tabular}{llcrlllcl}
\toprule
PN G &name & obs & obs date$^{(2)}$  &  &PN G & name & obs & obs date$^{(2)}$\\
\midrule
001.5$-$06.7 & SwSt1* & LCO & 08/09/09  & & 103.7+00.4 & M2-52* & SPM & 02/11/00\\
001.5$-$06.7 & ~~'' & SPM & 05/08/97  & & 104.4$-$01.6 & M2-53* & SPM & 02/11/00\\
002.2$-$09.4 & Cn1-5* & LCO & 09/09/09  & & 108.4$-$76.1 & BoBn1* & SPM & 27/08/01\\
002.2$-$09.4 &~~ '' & SPM & 17/06/96  & & 111.8$-$02.8 & Hb12* & SPM & 25/08/01\\
002.4+05.8 & NGC6369* & LCO & 05/06/10  & & 118.0$-$08.6 & Vy1-1* & SPM & 02/11/00\\
002.4+05.8 & ~~'' & SPM & 15/06/96  &  &118.8$-$74.7 & NGC246* & SPM & 13/12/98\\
003.1+02.9 & Hb4* & LCO & 05/06/10  &  &119.6$-$06.7 & Hu1-1* & SPM & 26/08/01\\
003.1+02.9 & ~~'' & SPM & 14/06/96  &  &120.0+09.8 & NGC40* & SPM & 13/12/98\\
003.9-14.9 & Hb7* & SPM & 26/08/01  & & 130.2+01.3 & IC1747* & SPM & 14/12/98\\
004.9+04.9 & M1-25 & LCO & 05/06/10  &  &130.3$-$11.7 & M1-1* & SPM & 02/11/00\\
004.9+04.9 & ~~'' & SPM & 17/06/96  &  &130.9$-$10.5 & NGC650-51* & SPM & 26/08/01\\
006.8+04.1 & M3-15* & LCO & 08/09/09  &  &133.1$-$08.6 & M1-2* & SPM & 27/08/01\\
006.8+04.1 & ~~''& SPM & 17/06/96  &  &144.5+06.5 & NGC1501 & SPM & 14/12/98\\
009.4$-$05.5 & NGC6629* & SPM & 05/08/97  &  &146.7+07.6 & M4-18* & SPM & 14/12/98\\
010.8$-$01.8 & NGC6578 & SPM & 14/06/96  &  &159.0$-$15.1 & IC351* & SPM & 05/10/99\\
011.7$-$00.6 & NGC6567* & SPM & 05/08/97  &  &161.2$-$14.8 & IC2003* & SPM & 05/10/99\\
011.9+04.2 & M1-32* & LCO & 05/06/10  &  &166.1+10.4 & IC2149* & SPM & 14/12/98\\
011.9+04.2 & ~~'' & SPM & 14/06/96  &  &194.2+02.5 & J900* & SPM & 02/11/00\\
012.2+04.9 & PM1-188 & SPM & 04/08/97  &  &197.8+17.3 & NGC2392* & SPM & 13/12/98\\
017.9$-$04.8 & M3-30* & SPM & 17/06/96  &  &243.3$-$01.0 & NGC2452* & SPM & 13/12/98\\
019.4$-$05.3 & M1-61 & LCO & 05/06/10  & & 278.1$-$05.9 & NGC2867 & LCO & 08/05/06\\
019.4$-$05.3 & ~~'' & SPM & 25/08/01  & & 278.8+04.9 & PB6 & LCO & 10/05/06\\
019.7$-$04.5 & M1-60 & SPM & 25/08/01  &  &285.4+01.5 & Pe1-1 & LCO & 04/06/10\\
025.8$-$17.9 & NGC6818* & SPM & 27/08/01  &  &286.3+02.8 & He2-55 & LCO & 10/05/06\\
027.6+04.2 & M2-43 & SPM & 17/06/96  &  &291.3$-$26.2 & Vo1 & LCO & 08/05/06\\
029.2$-$05.9 & NGC6751* & LCO & 09/09/09  &  &292.4+04.1 & PB8 & LCO & 09/05/06\\
037.7$-$34.5 & NGC7009* & SPM & 02/11/01  &  &294.1+43.6 & NGC4361* & SPM & 17/06/96\\
042.5$-$14.5 & NGC6852* & SPM & 27/08/01  &  &300.7$-$02.0 & He2-86 & LCO & 05/06/10\\
046.4$-$04.1 & NGC6803* & SPM & 26/08/01  &  & 307.2$-$03.4 & NGC5189* & LCO & 09/05/06\\
048.7+01.9 & He2-429 & SPM & 05/10/99  &  &309.0$-$04.2 & He2-99 & LCO & 09/05/06\\
051.9$-$03.8 & M1-73 & SPM & 26/08/01  &  &321.0+03.9 & He2-113 & LCO & 09/05/06\\
054.1$-$12.1 & NGC6891* & SPM & 25/08/01  &  &327.1$-$02.2 & He2-142 & LCO & 09/05/06\\
058.3$-$10.9 & IC4997* & SPM & 25/08/01  &  &332.9$-$09.9 & CPD-56 & LCO & 10/05/06\\
061.4$-$09.5 & NGC6905* & SPM & 14/06/96  &  &336.2$-$06.9 & PC14 & LCO & 05/06/10\\
064.7+05.0 & BD+30 3639* & SPM & 05/10/99  &  &337.4+01.6 & Pe1-7 & LCO & 04/06/10\\
086.5$-$08.8 & Hu1-2* & SPM & 25/08/01  &  &355.2$-$02.5 & H1-29 & SPM & 26/08/01\\
089.0+00.3 & NGC7026* & SPM & 14/12/98  &  &355.9$-$04.2 & M1-30* & LCO & 05/06/10\\
089.8$-$05.1 & IC5117* & SPM & 25/08/01  &  &355.9$-$04.2 & ~~'' & SPM & 27/08/01\\
096.3+02.3 & K3-61* & SPM & 05/10/99  &  &356.2$-$04.4 & Cn2-1*
 & SPM & 05/08/97\\
096.4+29.9 & NGC6543* & SPM & 15/06/96  &  &358.3$-$21.6 & IC1297 & LCO & 08/09/09\\
100.6$-$05.4 & IC5217* & SPM & 05/10/99  &    &   &\\
\bottomrule
\multicolumn{9}{l}{$^{(1)}$The observatory and observing dates are indicated.}\\
\multicolumn{9}{l}{~~~~Observations at LCO were obtained with MIKE, and at SPM, with the echelle REOSC.}\\
\multicolumn{9}{l}{~~~~Objects with * are part of the sample downloaded from the SPM Kinematical Catalogue.}\\
\multicolumn{9}{l}{$^{(2)}$Observing date in dd/mm/yy.}
\end{tabular}
\end{table*}

The data for LCO objects were obtained with the double echelle Magellan Inamori Kyocera  spectrograph, MIKE (Berstein et al. 2003). A full description of the observations and data reduction procedures  are presented by Garc{\'\i}a-Rojas et al. (2009) and Garc{\'\i}a-Rojas et al. (2012), who have used the data of  thirteen  of these objects to analyze the physical conditions and chemical behavior of the nebulae. In this case the spectral resolution is better than 0.17 \AA ~in the blue (about 10.8 km s$^{-1}$) and 0.23 in the red (about 12.8 km s$^{-1}$).

In addition, we have collected data from the SPM Kinematic Catalogue of Galactic Planetary Nebulae (L\'opez et al. 2012), which provides spatially resolved, long-slit echelle spectra for about 600 galactic PNe. Position-velocity  images in H$\alpha$, [NII]$\lambda\lambda$ 6548, 6583,  and [OIII]$\lambda$5007, obtained with different slit positions across the nebulae, are presented for each object.  From this catalogue, we downloaded  all the available spectra of [WR]PNe (slit passing through the center), in order  to measure  their systemic heliocentric radial velocity. In total we found 54 objects. Of these, 48  are in common with SPM and LCO  objects (they are marked with * in Table 1). The additional 6 objects, not observed at SPM or LCO, are PN G009.8-04.6, PN G068.3-02.7, PN G081.2-14.9, PN G189.1+19.8, PN G208.5+33.2,  and PN G307.5-04.9.

Our final sample, with measured radial velocities, consists of  78 objects of which 44 are [WR]PNe, 3 are [WC]-PG1159 objects (considered as one group in the following), and 16 PNe are ionized by wels (hereafter WLPNe). The remaining 15 PNe contain a normal or a  PG1159 central star. The [WR]PN sample represents the  43\% of the total sample of known [WR]PNe which presently amounts to 103 objects (De Pew et al. 2011). Medina et al. (2006) showed that,  regarding the expansion velocities,  WLPNe behave similarly to normal PNe. Gesicki et al. (2006) and other authors have found that there are noticeable differences between [WR]PNe and  WLPNe, so in the following, we will consider WLPNe and normal PNe in one group, apart from the [WR]PN group. This will be further discussed in \S 2.1 and \S 4.

Due to we are analyzing the distribution and galactic kinematics of [WR]PNe, we have included in our sample other 5 [WR]PNe (PN G020.9-01.1, PN G274.3+09.1, PN G306.422.4-00.1, PN G309.1-04.3, and PN G322.4-00.1) 
for which we found distances (given by Stanghellini \& Haywood 2010), but not velocities. They will be used for analyzing the galactic distribution of [WR]PNe. 

The main characteristics of all the analyzed objects are presented in Table 4 where we include, in  column 1,  the name corresponding to the Strasbourg-ESO Catalogue of Galactic Planetary Nebulae by Acker et al. (1992);   in  column 2,  the common name and, in column 3,  the  spectral classification of the central star: [WC\#] or [WO\#] for Wolf-Rayet central stars, `wels'  for weak emission line stars, and `pn' for normal stars. The [WR] or `wels' classification were adopted from   Acker \& Neiner (2003) and references therein, and Todt et al. (2010) for the case of PN G292.4+04.1 (PB\,8). Columns 4 and 5 of Table 4 show  the galactocentric distances of the objects and their errors, as given in the work by Stanghellini  \& Haywood (2010).

\begin{figure}[!h]
\begin{center}
  \includegraphics [scale=0.34]{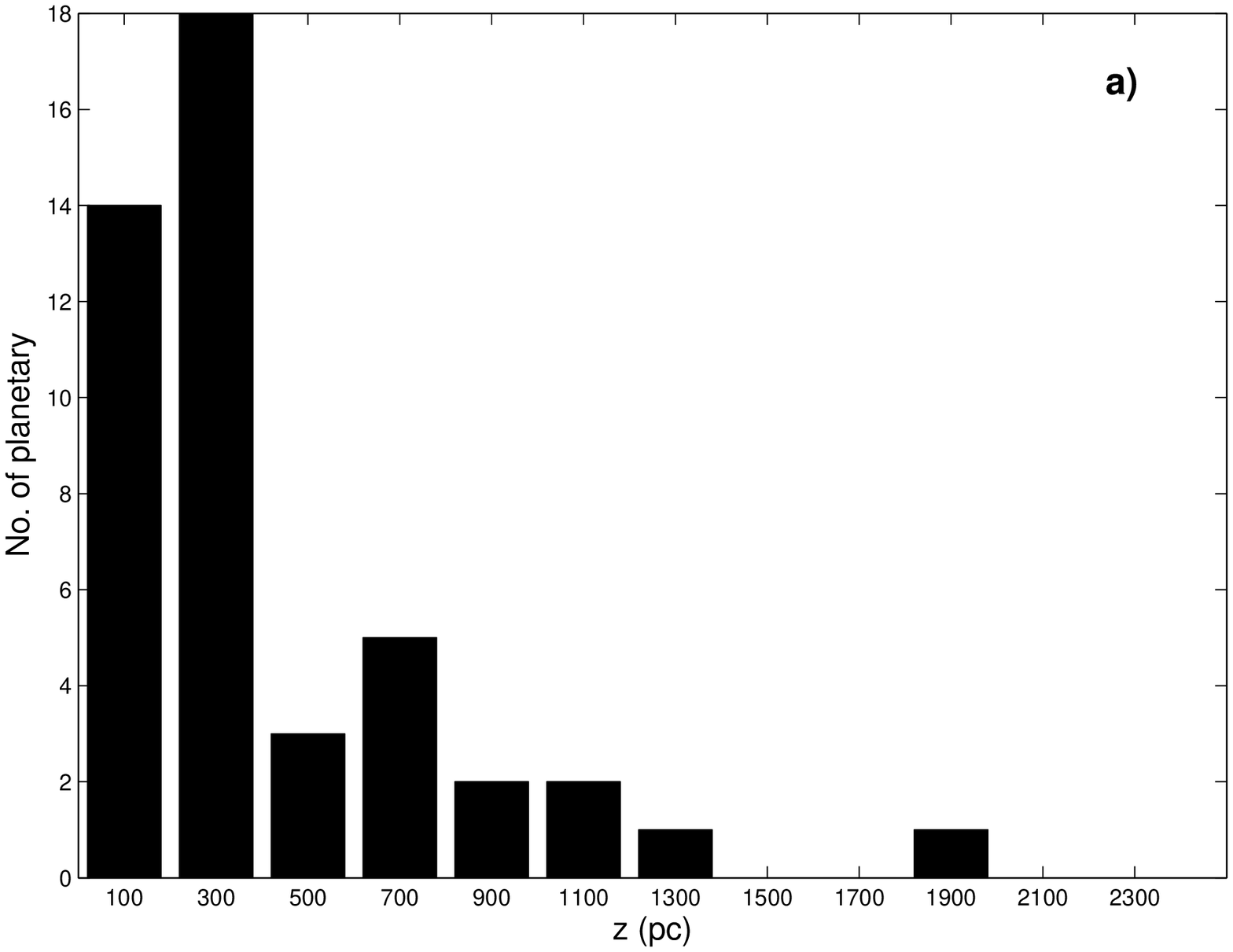}
  \includegraphics [scale=0.34]{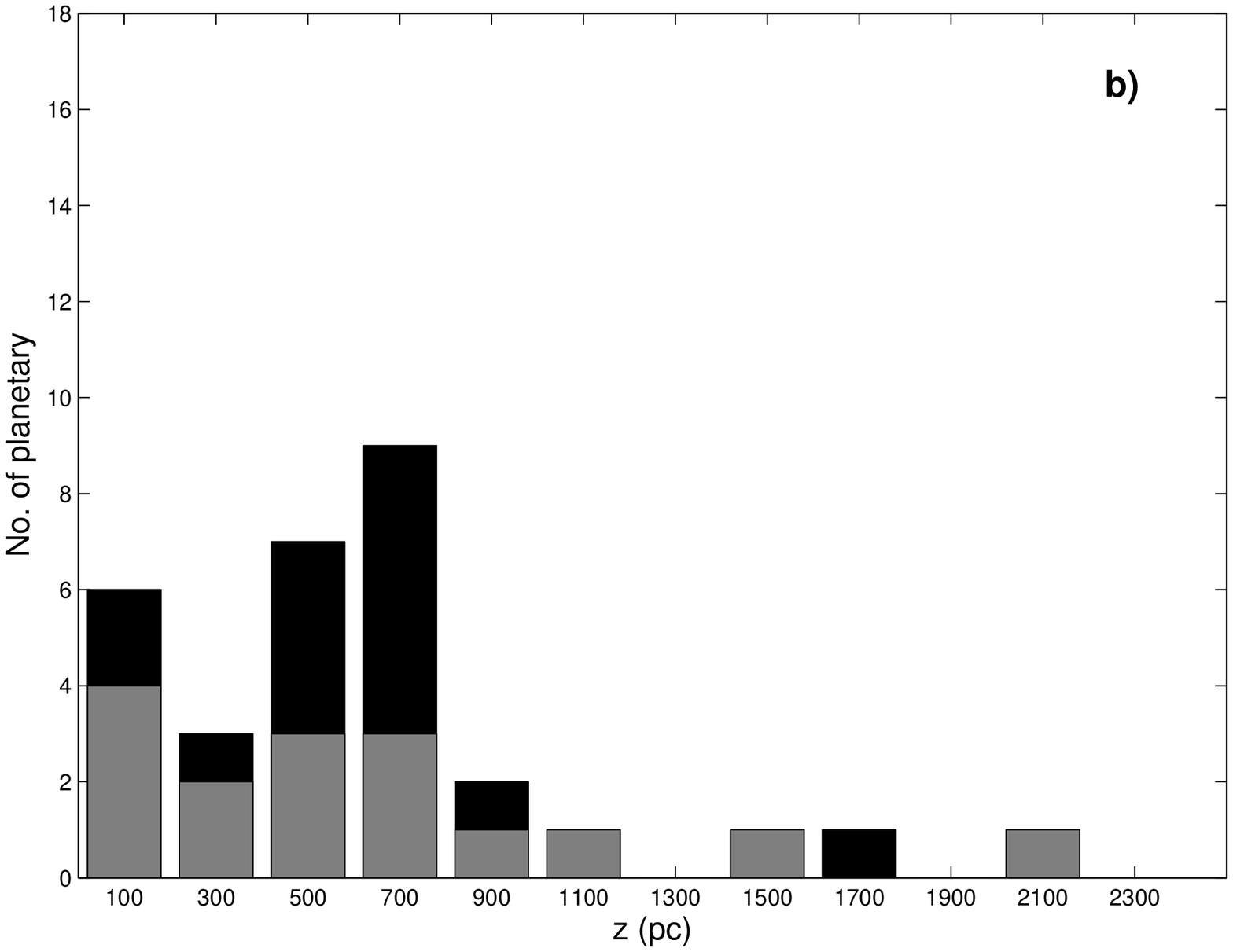}
  \caption{Distribution of objects at different height, $z$, above the galactic plane. (a) [WR]PNe (46 objects), (b)  objects ionized by wels (histogram in gray) and  by normal central stars (histogram in black); 31 objects in total. The halo PN G108.4-76.1(BoBn\,1) is not represented here. It has $z$ =  17 kpc}
    \end{center}
\end{figure}

\subsection{Distribution of [WR]PNe and WLPNe, relative to the galactic disk }

Figure 1 presents the distribution of objects  ([WR]PNe, WLPNe and normal PNe as a function of  height above the galactic disk, {\it z} (pc). Heights were  obtained by assuming the heliocentric distances given by Stanghellini \& Haywood (2010, distances were found for  78 objects)  and taking into account the galactic coordinates of the objects. The heights are listed in column 6 of Table 4. In the graph for [WR]PNe (Fig. 1a, which also includes three  [WC]-PG1159 stars) it is evident that most of the objects (32 of 46) belong to a thin disk with height lower than 400 pc, while in the graph for WLPNe  and normal PNe (Fig. 1b),  the great majority of the objects (25 of 31) presents height above the galactic plane up to a distance of 800 pc.  It is very interesting to notice that WLPNe  (gray histogram in Fig.1b) do not show any particular concentration towards the thin disk. Twelve of sixteen WLPNe have heights up to 800 pc and the other four objects are located at larger $z$, very similar to the distribution shown by PNe with normal central stars. Although there are only 16 WLPNe in our sample, this result is indicating that these objects  are distributed differently than [WR]PNe. 

Thus, regarding their position in the Galaxy, [WR]PNe seem to belong to a population located in  a thinner disk than PNe ionized by wels and normal central stars, indicating that progenitor stars of [WR]PNe would be younger and  with larger initial masses. In our [WR]PN sample there are only 4 objects with $z~ >$ 1 kpc; they are PN G358.3-21.6 (IC\,1297) with $z$= 1.827 kpc, PN G161.2-14.8 (IC\,2003) with $z$= 1.21 kpc, PN G146.7+0.6 (M\,4-18) with $z$=1.185 kpc, and PN G118.0$-$08.6 (Vy\,1-1) with $z$= 1.07 kpc.

 The galactic distribution of [WR]PNe was analyzed by Acker et al. (1996). They found that the fraction of [WR]PNe with galactic latitude $|b| < 7^o$  is similar to the fraction for normal PNe, thus concluding that both distributions are equal. However they did not considered the distances and  heights above the galactic plane, which is probably the reason for the difference with our results.

\section{Heliocentric, circular and peculiar radial velocities}
For  the objects observed at the OAN-SPM and LCO, radial velocities were measured from the most intense spectral lines (not saturated), such as  [OIII]$\lambda\lambda$ 5007, 4959, H$\gamma$, H$\beta$, H$\alpha$, [NII]$\lambda\lambda$6548, 6583, by using the task {\it splot} of IRAF{\footnote{IRAF is distributed by the National Optical Astronomy Observatories, which is operated by the Association of Universities for Research in Astronomy, Inc.  (AURA) under contract with the National Science Foundation.}. 

For the OAN-SPM data we found that the red zone of our spectra (wavelength longer than 6000 \AA) was not properly calibrated in wavelength, giving results with large discrepancies relative to the blue zone, therefore we used only the blue lines [OIII]$\lambda\lambda$ 5007, 4959, H$\gamma$, and H$\beta$ to determine the radial velocities of these objects. 
The data from LCO have  better  resolution and the blue and red lines show almost no discrepancy, then radial velocities were
measured for all the lines, (and their results are preferred over the OAN-SPM and SPM kinematical catalogue ones).

The finally adopted radial velocity in each case,  corresponds to the average obtained from  the used lines and the errors were calculated as the mean quadratic error, $\Delta$V = ($\Sigma_i(V_i - V)^2 / (n(n-1))^{1/2}$, where $V_i$ is the velocity for each line, V is the average velocity and n is the number of considered lines. Then, these errors  represent the internal consistency of our spectra and not necessarily the true uncertainty in the determined velocity (see \S 3.1).

 Afterwards we used the IRAF routine {\it rvcorrect} to determine, for each object, the heliocentric radial velocity and its error. The results are presented in columns 7 and 8 of Table 4. In these columns, the data from LCO are boldfaced. 
   
 As said before, we selected the galactic disk [WR]PNe, appearing in  the Kinematical Catalogue of SPM (L\'opez et al. 2012) and determined  the heliocentric radial velocities of these objects by  measuring the systemic velocity  of the nebula from the position-velocity diagram obtained when  the slit was positioned through the center of the object. We measured all the available lines for each object ([OIII]$\lambda$5007, H$\alpha$, and [NII]$\lambda\lambda$ 6548,6583). Such velocities, V$_{\rm Cat}$, are listed in column 9 of Table 4.  Notice that for a few objects this is the only available velocity.
  
\subsection{Analysis of velocities}
\begin{figure}[!th]
\begin{center}
  \includegraphics [scale=0.4]{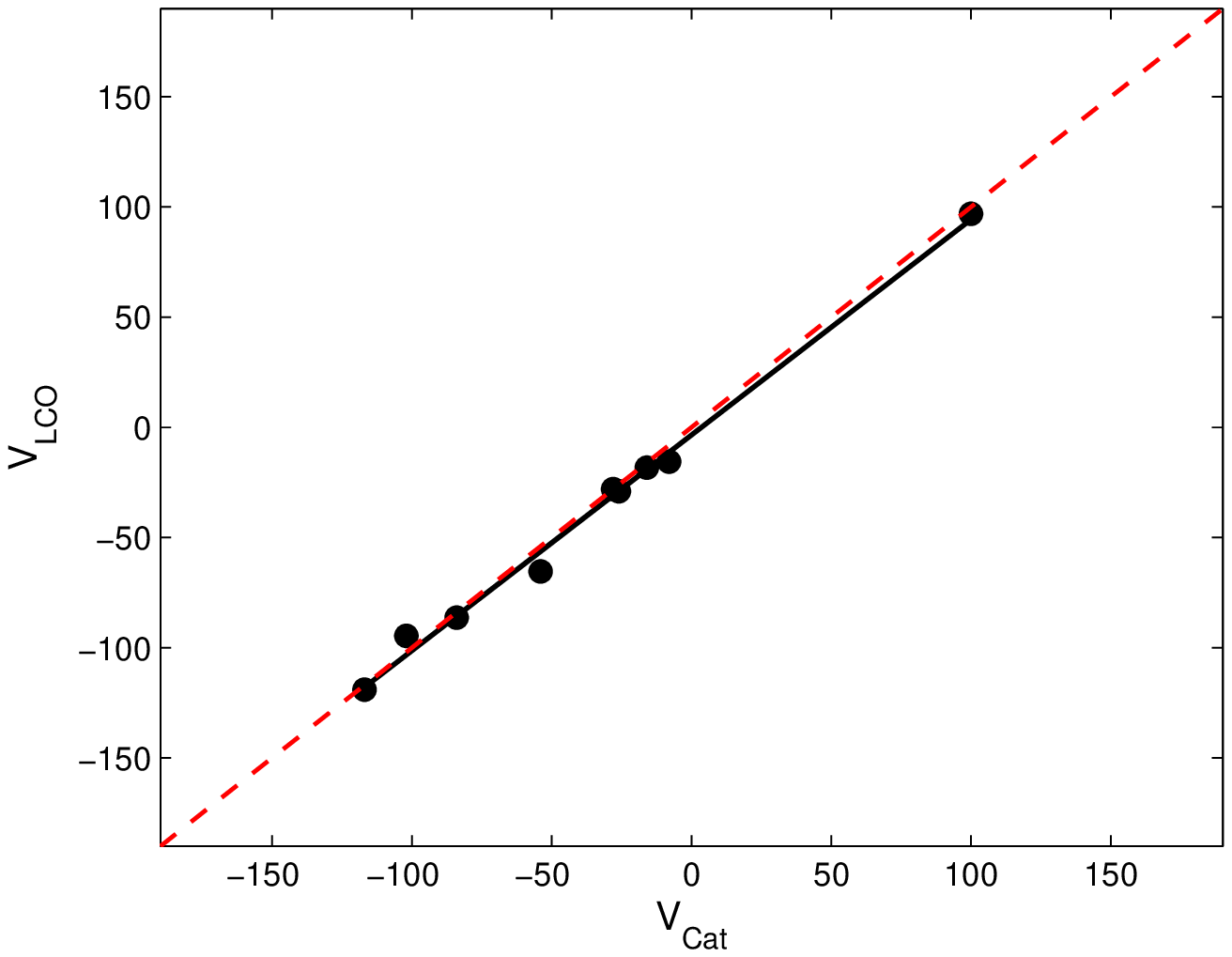}
  \includegraphics [scale=0.4]{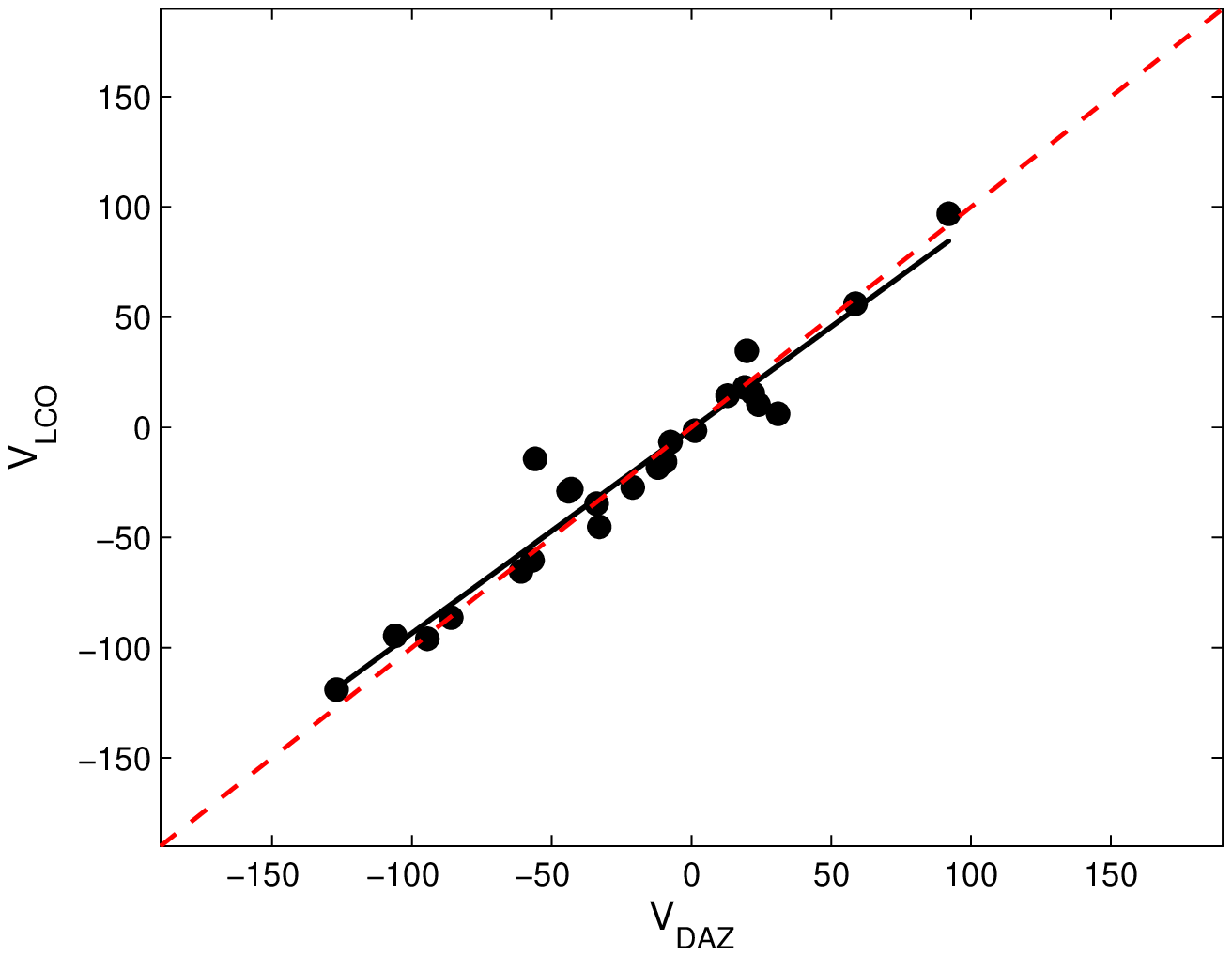}
 \includegraphics[scale=0.4]{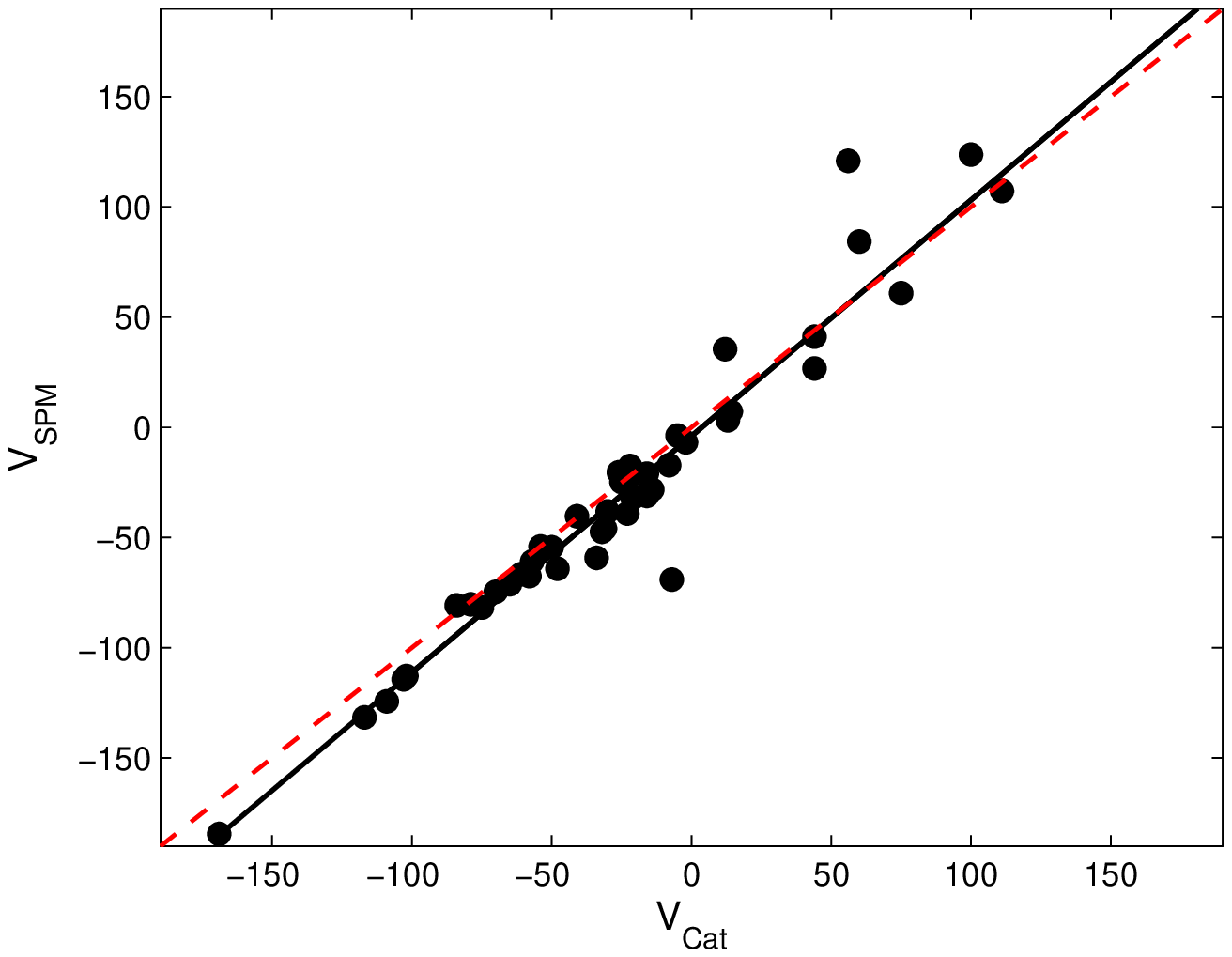}
  \includegraphics[scale=0.4]{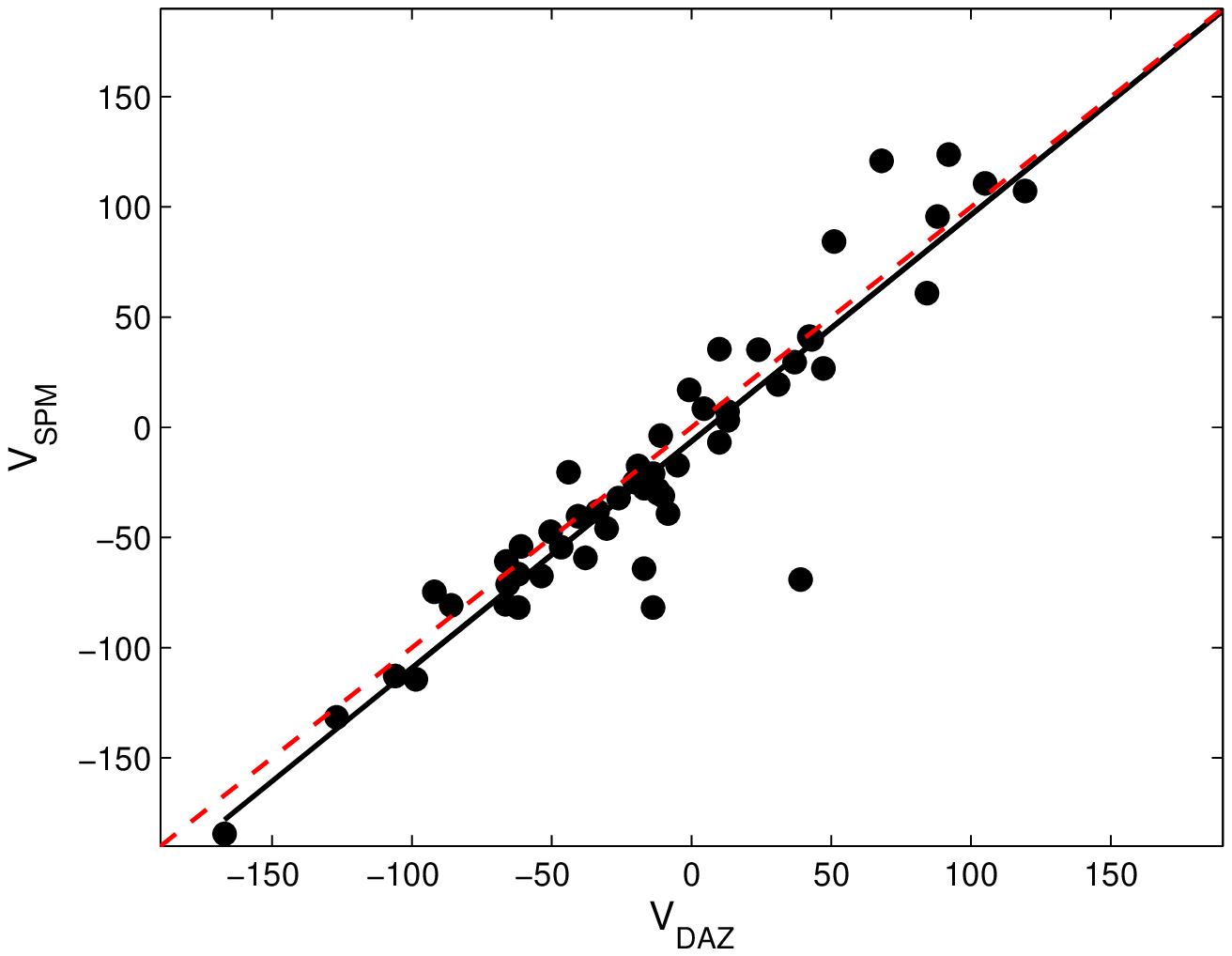}
     \caption{Comparison of heliocentric radial velocities from different observations. Top:  V$_{\rm LCO}$ vs. velocities from the SPM Kinematical Catalog, V$_{\rm Cat}$, and V$_{\rm LCO}$ vs. Durand et al. (DAZ, 1998)  data. Bottom: V$_{\rm SPM}$ vs. V$_{\rm Cat}$ and V$_{\rm SPM}$ vs. Durand et al. velocities. The solid  lines represent the linear fits to the data and they are discussed in the text. The dashes lines represent a 1:1 relation (45$^{\rm o}$ slope lines). }
    \end{center}
\end{figure}

An adequate  way of estimating the uncertainties in an observed quantity is by comparing independent measurements of the quantity. For the heliocentric velocities of our objects we have, in most of the cases, two or three independent observations that can be compared, and thus the uncertainties can be estimated.

Figure 2 shows a comparison of our heliocentric radial velocities (V$_{\rm LCO}$ and V$_{\rm SPM}$) with the values derived from the SPM Kinematical Catalogue, (V$_{\rm Cat}$), and with values from the literature (Durand et al. 1998, V$_{\rm DAZ}$).  A 45$^{\rm o}$ slope line has been included in all the graphs  for comparison. It is evident that  the velocities from LCO and from the Kinematical Catalogue (Fig. 2 up, left), which are the ones with the best spectral resolution, are very well correlated; the linear correlation is V$_{\rm LCO}$ = 0.978 V$_{\rm Cat} -$3.543 km s$^{-1}$, with a correlation coefficient r$^2$= 0.994. The  dispersion of the differences (V$_{\rm LCO}$ - V$_{\rm Cat}$) is 3.8 km s$^{-1}$. For the velocities from SPM we found a good correlation with those from the Kinematical Catalogue (Fig. 2 bottom, left), except for a few objects showing large differences. The linear fit is V$_{\rm SPM}$ = 1.072 V$_{\rm Cat} -$3.967, with r$^2$=0.948. The dispersion of the differences (V$_{\rm SPM}$ - V$_{\rm Cat}$) is  12.9 km s$^{-1}$.

The comparison of V$_{\rm LCO}$ vs. V$_{\rm DAZ}$ (Fig. 2 up, right) is also good with only one object  showing a discrepancy as large as 40 km s$^{-1}$.  The fit is V$_{\rm LCO}$ = 0.927V$_{\rm DAZ} -$0.719 with correlation coefficient r$^2$= 0.942. The dispersion of the differences (V$_{\rm LCO}$ - V$_{\rm DAZ}$)  is 8.8 km s$^{-1}$.   The correlation is worse for V$_{\rm SPM}$ vs. Durand et al. data (Fig. 2 bottom, right)  that has a fit V$_{\rm SPM}$ = 1.028 V$_{\rm DAZ} -$6.382, with correlation coefficient  r$^2$= 0.890.  The dispersion of the differences is, in this case, 18.0 km s$^{-1}$.
Durand et al. data correspond to a compilation  from the literature of results from different authors and, accordig to Durand et al.  most of their sample  presents velocity uncertainties better than about  20 km s$^{-1}$.  

 In all the graphs it is apparent that the 1:1 correlation and the linear fits are very similar, showing the good quality  of the different estimates of velocities. 

Considering the above discussion,  we have decided to adopt, when possible,  the velocities obtained from LCO  with an uncertainty of $\pm$3.8 km s$^{-1}$; if these are not available, we have adopted the velocities from the SPM Kinematical Catalogue with the same uncertainty and, as a third choice, we have used the velocities obtained with the Echelle REOSC from SPM, assuming an uncertainty of  $\pm$12.9 km s$^{-1}$.

\subsection{Circular velocities}

From  the galactocentric distance of the objects, R$_{\rm G}$, as given by Stanghellini \& Haywood (2010), we determined their circular  radial velocities  following the  expression: 

\begin{eqnarray}
 {\rm V}_{\rm circ}= -{\rm u_\sun \,  cos\,l~cos\,b -v_\sun \,sin\,l ~cos\,b } \nonumber \\  
 - {\rm w_\sun \,sin\,b -  2 A ( R_G -   R_\sun) \, sin\,l~cos\,b} \nonumber \\ 
 +  {\frac {\rm A_2} {2} } \, {\rm (R_ G}- {\rm R_\sun)^2~sin\,l ~cos\,b +K.}
\end{eqnarray}

\noindent This equation is the expansion to second order of the circular radial velocity as a function  of R$_{\rm G}$. The expansion to second order  allows to use the kinematics of objects located further from the Sun. In Eq. 1, u$_\sun$, v$_\sun$ and w$_\sun$ are the  radial, azimuthal and vertical components of the solar motion relative to the Standard Local of Rest.
R$_\sun$ is the distance of the Sun to the Galactic center and `l' and `b' are the galactic longitude and latitude of the object. 

The forth term in the right side of Eq. 1 represents the usual differential galactic rotation to first order, being A the Oort constant. In the second row,  A$_2$ is the second order coefficient of the derivative of the rotation speed with respect to R$_{\rm G}$, and  K is the K-term of local galactic expansion.

To calculate the radial circular velocities from Eq. 1, we used the standard values for the kinematical parameters that are listed in Table 2.
\setcounter{table}{1}
\begin{table}
\begin{center}
\caption{Kinematical parameters for Eq. 1}
\begin{tabular}{lll}
\toprule
parameter & value (units)& ref.$^{(1)}$\\
\midrule
$u_\odot$ & 10.4 (km s$^{-1}$) & 1\\
$v_\odot$ &  14.8 (km s$^{-1}$) & 1\\
$w_\odot$ &   7.3 (km s$^{-1}$) & 1\\
$R_\odot$ &   8.0$\pm$0.5 (kpc)& 2\\
$A$ &  14.4$\pm$1.2 (km s$^{-1}$ kpc$^{-1}$)& 3\\
$A_{2}$ &   13.0$\pm$0.9 (km s$^{-1}$ kpc$^{-2}$) & 4\\
$K$ & 5.1$\pm$2.8 (km s$^{-1}$) & 4\\
\bottomrule
\multicolumn{3}{l}{$^{(1)}$ \small 1. Mihalas \& Routly 1968; 2. Reid 1993;}\\
 \multicolumn{3}{l}{ \small 3. Kerr \& Lynden-Bell 1986; 4. Durand et al. 1998}\\ 
\end{tabular}
\end{center}
\end{table}

Values for the derived V$_{\rm circ}$ are listed in column 11 of Table 4. The errors for V$_{\rm circ}$ (columns 12 and 13) were computed by considering the errors in the galactocentric distances.

After calculating V$_{\rm circ}$ we determined the peculiar velocity for each object as:

\centerline{ ${\rm V_{pec} = V_{ hel} - V_{circ}}$.} 

The resulting values and the corresponding errors are listed in columns 14, 15, and 16 of Table 4. It is important to notice that the errors in distances are, in general, the most important uncertainties in computing V$_{\rm pec}$.

\section{Stellar populations of PNe}
\label{Description}

We have classified our sample of  PNe using the criteria of Peimbert 's classification scheme defined in the Introduction, thus Types I are the PNe with He/H$>$0.125, and N/O$>$0.5, Types II do not show  He or N enrichment and have V$_{\rm pec} \leq$ 60 km s$^{-1}$ and Types III have  V$_{\rm pec} >$ 60 km s$^{-1}$. The chemical composition for our objects (important for Type I classification) were adopted from the works by Garc{\'\i}a-Rojas et al. (2009; 2013 in preparation), Pe\~na  et al. (2001) or from the literature. For each Type I object  we have listed its relevant abundance ratios  (N/O and He/H) in Table 3 where the source for  abundances is identified.  

Peimbert's type for each object is shown in column 17 of Table 4. In Table 5 we present a summary of the number of [WR]PNe, WLPNe and normal PNe in each  Type. In next subsections a discussion on the different  Peimbert's type samples is given.

\begin{table*}[tH]
\begin{center}
\caption{Peimbert's Type I PNe ([WR]PNe are boldfaced).}
\label{modcel}
\begin{tabular}{lrrrrrrrrrr}
\toprule
 &object &V$_{\rm pec}$&  N/O & He/H & ref$^{(1)}$ \\
\midrule
{\bf 001.5$-$06.7}& {\bf SwSt\,1}&$-$17.0 &0.80 &0.040& PSM01\\
{\bf 002.2$-$09.4}&\textbf{Cn\,1-5}&  $-$37.9 &  0.83 & 0.158&G-R12\\
{\bf 003.1+02.9}&\textbf{Hb\,4}&-75.9&   0.71& 0.115& G-R12\\
{\bf 011.9+04.2}&\textbf{M\,1-32}&$-$132.3 & 0.51& 0.126& G-R12\\
{\bf 064.7+05.0}&\textbf{BD+30 3036}&$-$40.5 &1.25& ---& PSM01\\
086.5$-$08.8&Hu\,1-2&19.0 & 1.15& 0.151 & PLT-P95 \\
{\bf 089.0+00.3}&\textbf{NGC\,7026}&$-$24.8 & 1.01 & 0.124& PSM01\\
103.7+00.4&M\,2-52&$-$32.3 & 2.3 & 0.165& PM02\\
{\bf 278.8+04.9}&\textbf{PB\,6}&25.1 &  1.48 & 0.19 &G-R09\\
{\bf 300.7$-$02.0}&\textbf{He\,2-86}&5.9 &0.72&0.123 & G-R12 \\
{\bf 307.2$-$03.4}&\textbf{NGC\,5189}&$-$16.9 & 0.68 & 0.123 & KB94\\
{\bf 307.5$-$04.9} & \textbf{MyCn\,18} &  $-$34.6  & 1.04 & 0.095 & KB94\\
\bottomrule
\multicolumn{6}{l}{\small {$^{(1)}$ PSM01: Pe\~na et al. 2001; G-R12: Garc{\'\i}a-Rojas et al. 2013 in prep;}}\\
\multicolumn{6}{l}{\hskip 0.45cm {\small  PLT-P95: Peimbert et al. 1995; PM02: Pe\~na \& Medina 2001;}}\\
\multicolumn{6}{l}{\hskip 0.45cm {\small G-R09: Garc{\'\i}a-Rojas et al. 2009; KB94: Kingsburgh \& Barlow 1994}}\\ 
\end{tabular}
\end{center}
\end{table*}

\subsection{Type I objects}
As expected, Type I PNe show, in general, low peculiar velocities (lower than 50 km s$^{-1}$). This  agrees with these kind of PNe being supposedly produced by the most massive stars among progenitors (M$_{\rm i} \geq $ 2.0 M$_\sun$, Peimbert \& Serrano 1980). 

In our [WR]PN+[WR]-PG1159 sample, there are 10 Type I PNe and among  them there are two objects with high peculiar velocities: PN G003.1+02.9 (Hb\,4) with V$_{\rm pec} = -$75.9 km s$^{-1}$, and PN G011.9+04.2 (M\,1-32), with V$_{\rm pec}=-$132.2. If we follow the standard criteria for considering an object as belonging to the bulge, which are: galactic position within 10$^{\rm o}$ from the galactic center, radius lower than 20$''$ and radio flux at 5 Ghz  smaller than 100 mJy (Stasi\'nska \& Tylenda 1994;  Cavichia et al. 2011), Hb\,4  could be considered as  belonging to the bulge, because it is located within 10$^{\rm o}$ from the galactic center, it has a radius of 2.5$''$  and its flux at 5 Ghz (6 cm)  is 166 mJy. Then, Hb\,4 fulfills all the criteria except that it is  brighter at 5 Ghz. Its galactocentric distance is about 2.9$\pm$1.0 kpc, so it could be part of the bulge, despite its relatively strong 5 Ghz flux.

M\,1-32 is a peculiar object. Its chemical abundances (N/O= 0.51 and He/H = 0.126) locate it marginally among Type I PNe. If Kingsburgh \& Barlow (1994) criteria were adopted, it would not  be a Type I PN and its peculiar velocity would locate it among Type III's despite its height above the galactic plane is only 0.35 kpc.
The internal kinematics of M\,1-32 is peculiar; Medina et al. (2006) found high velocity wings in the nebular lines, which are also apparent in Fig. 1 of Garc{\'\i}a-Rojas et al. (2012). Recently Akras \& L\'opez (2012) analyzed the velocity field in this object and confirmed the high velocity wings, which they interpret as due to collimated bipolar outflows reaching velocities of $\pm$200 km s$^{-1}$. These authors consider   M\,1-32 as a bulge object. Its angular diameter is 7.6$''$ and its observed flux at 5 Ghz is 61 mJy, so these two criteria are fulfilled for being a bulge PN. However it is located at more than 10$^{\rm o}$ from the galactic center, at a galactocentric distance of about 3.46$\pm$0.84 kpc.  With these characteristics M\,1-32 could be at the border of the bulge and its high V$_{\rm pec}$ could be a consequence of its location in the Galaxy.

Regarding WLPNe, it is important to remark that none of the objects in our sample is a Type I PN. This fact was already mentioned by Pe\~na et al. (2003) for a smaller sample, and also by Fogel et al. (2003), who analyzed a sample of 42 WLPNe finding none Type I among them.

Our sample includes only a small number (12) of normal PNe chosen randomly (11 objects belong to the disk and one is in the galactic halo), which is by no means representative of the total galactic PNe. In this short sample we have 2 Type I PNe (17\%), a fraction similar to what is found in larger samples (e.g., Peimbert  1990).

\subsection{Type II objects}
Among the [WR]PNe and [WC]-PG1159 PNe, there are 23 objects (51\% of the sample)  classified as Type II. This also occurs in normal PNe, where  the Type II objects are the majority (see e.g., Peimbert 1990; Stanghellini \& Haywood 2010). Among our normal PN and WLPN samples, there are 6 and 7 Type II PNe, respectively. These PNe belong to the disk intermediate population, their progenitors had initial masses lower than 2 M$_\odot$, and in consequence they would be older than Types I. 

\subsection{Type III objects}
Interestingly, there are five [WR]PNe among the Type III objects which, as we said, have V$_{\rm pec}$ larger than 60 km s$^{-1}$ and are the older PNe among the disk population, belonging probably to the thick disk. The 5  objects  represent the 11.6\% of our  [WR] sample; they are PNG 002.4+5.8 (NGC\,6369, [WC]4, V$_{\rm pec}=-89.6$ km s$^{-1}$), PN G029.3$-$05.9 (NGC\,6751, [WC]4, V$_{\rm pec}=-62.5$ km s$^{-1}$), PN G096.3+02.3 (K\,3-61 [WC]4-5, V$_{\rm pec}=-79.1$ km s$^{-1}$), PN G146.7+07.6 (M\,4-18, [WC]11, V$_{\rm pec}=-160.8$ km s$^{-1}$),  and PNG 336.2$-$06.9 (PC\,14 [WO]4, V$_{\rm pec}=-69.2$ km s$^{-1}$).   Finding Type III PNe among the [WR]PNe is peculiar because in \S 2.1 we have shown that  [WR]PNe are in general closer to the galactic plane than normal PNe and this would indicate that [WR]PNe are young objects. Among the few objects  possessing {\it z} larger than 800 kpc (see Fig. 1)  only PN G146.7+07.6 (with {\it z}=1.19 kpc) belongs to the Type III group.

One plausible explanation for these unexpected Type III [WR]PNe could be the errors in the adopted distances. For the objects with V$_{\rm pec}$ slightly above 60 km s$^{-1}$, the large V$_{\rm pec}$ error bars can move them into  Type II's. However there are at least three objects that would remain as genuine Type III:  PN G002.4+5.8, PN G146.7+07.6, and PN G336.2$-$06.9. The first one, NGC\,6369,  is a well-known extended  nearby PN. Stanghellini \& Haywood (2010) attribute it a heliocentric distance of 1.089 kpc,  very similar to the distance given by Zhang (1995)  of 0.92 kpc (with Zhang's distance we derive V$_{\rm pec}$=$-$89.1 km s$^{-1}$), therefore, the errors in the distance do not seem to be an explanation. Possibly the central star of NGC\,6369 is an old low-mass star belonging to the thick disk, despite its [WR] condition and its low height above the galactic plane.

For the case of PN G146.7+07.6 (M\,4-18) Stanghellini \& Haywood (2010) locate it at a heliocentric distance  of 8.96 kpc and at a galactocentric distance of 16.17 kpc, therefore  we derived a  peculiar velocity of $-$160.8 km s$^{-1}$. If the heliocentric distance given by Zhang (1995) of 6.85 kpc is adopted, the galactocentric distance is 14.2 kpc, and the peculiar velocity results to be $-$91.5 km s$^{-1}$, that is still high, classifying M\,4-18 as a Type III PN anyway. The distance given by Zhang (1995) is similar to the one derived independently by De Marco \& Crowther (1999). In order to choose the  more appropriate distance we can calculate the H$\beta$ luminosity of this object, by assuming the observed log(F(H$\beta$))= $-$11.89 (Acker et al. 1992) and the different heliocentric distances. When the distance by Stanghellini \& Haywood is used, we get log (L(H$\beta$)/L$_\odot$) = 0.50 that  is slightly high for a PN (see Fig. 5a by Pe\~na et al. 2007), while  using the distance by Zhang (1995) produces a log(L(H$\beta$)/L$_\odot$) = 0.27, typical of a not too-bright PN. Thus, we consider that the distance by Zhang (1995) is more adequate for  this [WR]PN, although it  is  still a Type III object and its central star would be a low mass object belonging to the thick disk.

For the case of PN G336.2$-$06.9 (PC\,14), Stanghellini \& Haywood heliocentric distance is 6.147 kpc, with a galactocentric distance of  3.451 kpc. This provides a V$_{\rm pec}$= 69.2 km s$^{-1}$ . If the distance by Zhang (1995)  of 5.11 kpc is used,  V$_{\rm pec}$ results to be 
51.3 km s$^{-1}$  and the object would be a Type II.

Therefore in our [WR]PN sample, only NGC\,6369 and M\,4-18 would be bonafide Type III PNe.

 Regarding non-[WR]PNe, we found three  normal and seven WLPNe among Type III PNe which is an adequate number for normal objects and a little too high for WLPNe. As our numbers are small, in this case our results are only tentative.

\subsection{The bulge objects}
Apart from the two possibly bulge objects discussed in \S 4.1, there are several PNe in our sample, for which we have 
determined heliocentric radial  velocities, that have been classified as bulge PNe by some authors (G\'orny et al. 2009; Cavichia et al. 2011). As these objects  do not rotate with the inner disk, it has not much sense to calculate their circular velocities. We have done it for completeness.  Although the number of bulge objects in our sample is very small, it is interesting to analyze their heliocentric  velocities in comparison with other samples of bulge PNe.  We have PN G004.9+04.9 (M\,1-25) with V$_{\rm hel} = 10.3$ km s$^{-1}$, PN G006.8+04.1 (M\,3-15) with  V$_{\rm hel}= 96.9$ km s$^{-1}$, PN G009.8$-$04.6 (H\,1-67) with V$_{\rm hel}= -15.0$ km s$^{-1}$) , PN G355.2$-$02.5 (H\,1-29) with V$_{\rm hel}=  -$27.7km s$^{-1}$), PN G355.9$-$04.2 (M\,1-30), with V$_{\rm hel}= -$117.0 km s$^{-1}$, and PN G356.2$-$04.4 (Cn\,2-1) with V$_{\rm hel}=-169.0$ km s$^{-1}$).  The heliocentric  velocities of these objects are mainly positive when the galactic longitude $l$ is between 0$^o$ and 10$^{\rm o}$, and negative when $l$ is from 0$^{\rm o}$ to -10$^{\rm o}$.  
In this sense they behave similarly to the sample discussed by Durand et al. (1998, see their Fig. 6), who have interpreted this behavior as rotation of the bulge.

\subsection{ Special cases}

Among our PNe there are two objects with a very large V$_{\rm pec}$. 
One is  PN G108.4-76.1 (BoBn\,1), a very interesting PN with a strange chemical composition as it shows Ne/O larger than 1 (Pe\~na et al. 1993; Otsuka et al. 2010),  identified as belonging to the galactic halo at a very large height above the plane (17 kpc). Zijlstra et al. (2006) have argued that this object could be part of the Sagittarius Dwarf Spheroidal galaxy, which is supported by the PN galactic position and its large V$_{\rm pec}$. 

Other PN with extremely high V$_{\rm pec}$ ($-$412.0 km s$^{-1}$) is PN G111.8$-$02.8 (Hb\,12)   but  this could be  due to a possibly  erroneous heliocentric distance of 14.25 kpc (galactocentric distance  18.47 kpc) given by Stanghellini \& Haywood (2010). The   heliocentric distance of 8.11 kpc, given by  Zhang (1995), locates the object much closer  and its galactocentric  distance results to be 13.34 kpc. With this, we derive a much more reasonable V$_{\rm pec}$ of $-$32.9 km s$^{-1}$. In addition, if we calculate the total H$\beta$ luminosity for this object (taking the apparent H$\beta$ flux by Acker et al. 1992), log(F(H$\beta$)) = 10.98), for a distance of 14.25 kpc, we get log(L(H$\beta$)/L$_\odot$) = 1.82, which is too high for a PN and  more typical of a compact HII region (see again Fig. 5a by Pe\~na et al. 2007).  When assuming the distance by Zhang (1995), we obtain  Log(L(H$\beta$)/L$_\odot$) = 1.33, which is an adequate H$\beta$ luminosity for a bright PN. Therefore we consider than the distance by Stanghellini \& Haywood (2010) might be an errata.

\setcounter{table}{4}
\begin{table}[tH]
\begin{center}
\caption{Distribution of objects in Peimbert's types}
\label{types}
\begin{tabular}{lcccc}
\toprule
 &Ty I& Ty II & Ty III & bulge  \\
\midrule
{[WR]}PN$^{(1)}$ & 10 & 23 & 5 & 5  \\
WLPN & 0 & 7 & 7 & 2\\
PN & 2 & 6 &3 & -- \\
\bottomrule
\multicolumn{5}{l}{\small {$^{(1)}$ [WR]PNe and [WR]-PG1159 included.}}\\
\multicolumn{5}{l}{\small {$^{(1)}$ The  9 missing [WR]PNe do not have velocities }}\\
\multicolumn{5}{l}{\small \hskip 0.4cm  or distances.}\\
\end{tabular}
\end{center}
\end{table}

\section{Conclusions}
-- From high-quality high-spectral resolution spectra we have determined the heliocentric radial velocities of a sample of [WR]PNe (43 objects, representing the 42\% of the total known sample), a sample of WLPNe  and PNe with normal central star. These data, together with distances obtained from the literature, allowed us to determine  the galactic kinematics of the objects. The radial circular and peculiar velocities were computed.  We found that these quantities are largely affected by errors in the assumed distances and 
for some objects where too large V$_{\rm pec}$ are obtained, this can be attributed to poorly determined distances.

-- We have found that most of the analyzed [WR]PNe are located in the galactic disk and they are more concentrated towards the thin disk (height smaller than about 400 pc from the disk) than the WLPNe and PNe with normal central star, most of which are distributed up to 800 pc from the disk. 

-- According to their chemical composition and peculiar velocities we have classified the studied sample in Peimbert's Types.  Nine [WR]PNe (21\%)  and one [WR]-PG1159 PN have been classified as Peimbert's Type I (they are N- and He-rich) and would  have progenitor stars with initial masses  larger than 2 M$_\odot$. Therefore they would be young objects (ages between 0.1 to 1 Gyr).  This is confirmed by their kinematics as all of them (except two objects of the bulge) show V$_{\rm pec}$ smaller than 50 km s$^{-1}$. 

-- The [WR]PNe with V$_{\rm pec} \leq$ 60 km s$^{-1}$ and no particular He or N enrichment, amounting 23 objects (51\% of the sample),  are catalogued as Peimbert's  Type II. They belong to the disk population and would be of intermediate age.  Interestingly, there are a five [WR]PNe with V$_{\rm pec}$ larger than 60 km s$^{-1}$ which are classified as  Type III PNe. Although some of them could be of Type II by considering the uncertainties in their distances, there are two objects which appear as genuine Type III [WR]PNe. With this classification, these  objects would be of the thick disk with an old  low-mass central star, indicating that the [WR] phenomenon can occur also in less massive and old progenitors.

-- In our sample of 16 WLPNe, no one is  a Peimbert Type I, result also found by other authors in the analysis of larger samples (e.g., Fogel et al. 2003). Our WLPNe are distributed in 7 Types II, 7 Types II and 2 bulge objects. With these characteristics WLPNe are objects belonging to the intermediate and old disk population, with progenitors of low initial masses. Thus it is  corroborated that  [WR]PNe and WLPNe are unrelated objects.

-- We have obtained the radial velocity of the [WR] PN G332.9$-$09.9 (He3-1333 whose central star, CPD-56 8032, has a [WC]10 spectral type) that has not been reported previously. Its heliocentric distance is not reported in the literature, therefore its peculiar velocity can not be calculated.

\acknowledgments
J.S. R.-G. acknowledges scholarship from CONACyT- M\'exico. J. G.-R. received financial support from the spanish Ministerio de Educaci\'on y Ciencia (MEC), under project AYA2007-63030. This work received financial support from DGAPA-UNAM (PAPIIT project IN105511).

\end{document}